\documentclass[%
 reprint,floatfix,
superscriptaddress,
 amsmath,amssymb,
 aps,
prl,
]{revtex4-2}

\usepackage{graphicx}
\usepackage{dcolumn}
\usepackage{bm}
\usepackage{color}
\usepackage{physics}
\usepackage{tensor}
\usepackage{mathrsfs}
\usepackage{hyperref}

\usepackage[dvipsnames]{xcolor}

\begin{document}


\title{Evolution of Cosmic String Loops under Gravitational Backreaction}


\author{Lasse Gerblich}

\affiliation{%
 Centre for Theoretical Cosmology, Department of Applied Mathematics and Theoretical Physics,
University of Cambridge, Wilberforce Road, Cambridge CB3 0WA, United Kingdom
}%


\author{Richard A. Battye}
\affiliation{
Jodrell Bank Centre for Astrophysics, Department of Physics and Astronomy, University of Manchester, Manchester, United Kingdom
}%

\author{E. Paul S. Shellard}%
\affiliation{%
 Centre for Theoretical Cosmology, Department of Applied Mathematics and Theoretical Physics,
University of Cambridge, Wilberforce Road, Cambridge CB3 0WA, United Kingdom
}%


\date{\today}

\begin{abstract}
Nambu-Goto cosmic string loops generically develop cusps, points where the string momentarily reaches the speed of light. These cusps produce strong gravitational wave bursts with a characteristic strain spectrum $\tilde{h}(\omega)\propto (G\mu)\,\omega^{-4/3}$, making them prime targets for gravitational wave searches, where $\mu$ is the string mass per unit length. However, this picture is modified when one accounts for gravitational backreaction. Using a convenient gauge, we reformulate the Nambu-Goto equations of motion for a loop moving in its own dynamically sourced gravitational field, enabling the first continuous numerical evolution of loops under this backreaction. A key finding is that locally cusps survive backreaction. Nevertheless, the gravitational waveform calculated in this formalism is significantly modified, weakening the cusp burst and introducing a high-frequency cutoff at  $f_c\propto(G\mu)^{-3/2}$. This suppression above $f_c$  reduces the expected signal-to-noise ratio in current and near-future detectors relative to unperturbed waveform predictions.
\end{abstract}

\maketitle

\textit{Introduction}---Cosmic strings are one-dimensional topological defects that form in the early Universe during symmetry-breaking phase transitions with a non-simply connected vacuum manifold~\cite{TWBKibble_1976,Hindmarsh_1995,Vilenkin:2000jqa}. They are predicted across a wide range of Beyond Standard Model theories, such as axion models~\cite{Vilenkin:2000jqa}, Grand Unified Theories~\cite{Dunsky_2021} and string-theoretic scenarios~\cite{Copeland_2010}, and constitute one of the most generic and robust signatures of high-energy symmetry breaking in the early Universe. 

If the symmetry being broken is a local gauge-symmetry the string energy is concentrated very tightly around a narrow string core and interactions of the fields that make up the string are exponentially suppressed outside of the core. Therefore, these strings can be treated as one-dimensional Nambu-Goto strings that only interact gravitationally with the surrounding spacetime. This picture is also applicable to fundamental superstrings of cosmological length~\cite{Vilenkin:2000jqa,Copeland_2010}. The dynamics of Nambu-Goto string loops in flat space admits a striking generic feature: \textit{cusps}, points on the string that momentarily travel at the speed of light. At a cusp the string emits a powerful, narrow beam of gravitational radiation with a characteristic strain spectrum $\tilde{h}(f)\propto f^{-4/3}$~\cite{Burden_1985,Vachaspati_1985,Damour_2000,Damour_2001,Damour_2005}. Such cusp bursts would lie within the sensitivity bands of current ground-based detectors~\cite{Abbott_2018} and of the planned LISA space interferometer~\cite{LISAWaveforms,Dimitriou_2025}, making them a prime observational target for cosmic string searches.

The string's mass generates a gravitational perturbation of the surrounding spacetime, which in turn acts back on the string itself. This \textit{gravitational backreaction} modifies the loop's trajectory and energy distribution and ultimately governs the long-term decay of the loop~\cite{PhysRevD.42.2505,Blanco_Pillado_2019,Wachter_2017,Wachter_2024}. 

In this paper we present a continuous-time, semi-perturbative formalism to model the evolution of cosmic string loops under gravitational backreaction, using the temporal-transverse gauge. Using this formalism we can resolve the intra-period dynamics of the loop and properly study the fate of a cusp. For the simple cosmic string loops we consider we find that locally cusps survive backreaction.

Our formalism further permits, for the first time, the direct computation of the gravitational wave waveform produced by a continuously backreacted Nambu-Goto string. We see that while the cusp survives locally on the string worldsheet the backreaction nevertheless affects the gravitational wave signal: the sharp cusp peak in the waveform is smoothed, and the power spectrum acquires a high-frequency cut-off whose scaling we determine numerically and motivate analytically using recent results in~\cite{drew2025newmasslessspectracosmic}. These findings have a significant impact on the detectability of cusp bursts with LIGO and LISA.

We work in natural units with $c=1$, and adopt the mostly plus $(-,+,+,+)$ metric signature for the spacetime, and analogously the $(-,+)$ signature for the worldsheet metric. Greek indices run over spacetime coordinates $\mu=0,1,2,3$, while Latin indices from the beginning of the alphabet run over worldsheet coordinates $a=0,1$.

\textit{Nambu-Goto Dynamics}---In the thin-string limit we regard the string as truly one-dimensional, so it traces out a two-dimensional worldsheet $\mathcal{W}$ described by $X^\mu(\zeta^a)$, where $\zeta^a, \; a=0,1$ parametrise the worldsheet. The dynamics of such a string are described by the Nambu-Goto action \cite{Vilenkin:2000jqa}
\begin{eqnarray}
    S=-\mu\int_\mathcal{W} \dd^2{\zeta} \: \sqrt{-\gamma},
    \label{eq:Nambu-Goto-Action}
\end{eqnarray}
where $\mu$ is the string tension and $\gamma=\det(\gamma_{ab})$ is the determinant of the induced metric on the worldsheet $\gamma_{ab}=g_{\mu\nu}\partial_a X^\mu \partial_b X^\nu$. Varying \eqref{eq:Nambu-Goto-Action} with respect to $X^\mu$ yields the Nambu-Goto equation of motion
\begin{eqnarray}
    \frac{1}{\sqrt{-\gamma}}\partial_a\left(\sqrt{-\gamma}\gamma^{ab}\partial_b X^\rho\right)+P^{\mu\nu}\Gamma^\rho_{\mu\nu}=0,
    \label{eq:Nambu-Goto-EoM}
\end{eqnarray}
with $\Gamma^\rho_{\mu\nu}$ the Christoffel symbols of the spacetime, $\gamma^{ab}$ such that $\gamma^{ac}\gamma_{cb}=\delta^a_b$ and $P^{\mu\nu}=\gamma^{ab}\partial_a X^\mu \partial_b X^\nu$~\cite{carter2001essentialsclassicalbranedynamics}.\\
Similarly, if we vary \eqref{eq:Nambu-Goto-Action} with respect to $g_{\mu\nu}$ we obtain the stress-energy tensor of the string
\begin{eqnarray}
    T^{\mu\nu}(x)=\frac{-\mu}{\sqrt{-g}}\int_\mathcal{W}\dd^2{\zeta}\:\sqrt{-\gamma}\:P^{\mu\nu}\delta^{(4)}\left(x-X(\zeta)\right),
    \label{eq:Stress-Energy-Tensor}
\end{eqnarray}
In the following, we will parametrise the worldsheet such that we identify one of the worldsheet coordinates with the coordinate time $t$ of the spacetime and furthermore choose the remaining coordinate $\zeta$ such that $\gamma_{t\zeta} = g_{\mu\nu}\dot{X}^\mu\acute{X}^\nu=0$, where $\dot{}$ and $\acute{}$ denote derivatives with respect to $t$ and $\zeta$, respectively.\\
We can then rewrite the worldsheet metric as
\begin{eqnarray}
    \gamma_{ab} = \mqty(\gamma_{tt} & 0 \\ 0 & \gamma_{\zeta\zeta})=\mqty(-\frac{\phi}{\epsilon} & 0 \\ 0 & \epsilon\phi),
\end{eqnarray}
with $\phi=\sqrt{-\gamma_{tt}\gamma_{\zeta\zeta}}$ and $\epsilon=\sqrt{-\frac{\gamma_{\zeta\zeta}}{\gamma_{tt}}}$. We can think of $\epsilon$ as the linear energy density of the string. This can be shown by applying the temporal-transverse gauge conditions to the stress energy tensor \eqref{eq:Stress-Energy-Tensor} and computing the total energy of the string loop
\begin{eqnarray}
    E(t)=\mu\oint\dd{\zeta}\: \epsilon(t,\zeta).
\end{eqnarray}
Applying the temporal-transverse gauge conditions to \eqref{eq:Nambu-Goto-EoM} we obtain the equation of motion
\begin{align}
\begin{split}
    &\dv{}{t}\left(\epsilon\dot{X}^\rho\right)-\dv{}{\zeta}\left(\frac{\acute{X}^\rho}{\epsilon}\right)\\
    &\quad=\left(-\epsilon\dot{X}^\mu\dot{X}^\nu+\frac{1}{\epsilon}\acute{X}^\mu\acute{X}^\nu\right)\Gamma^\rho_{\mu\nu}.\label{eq:TempOrtho-EoM}
\end{split}
\end{align}

\textit{Gravitational Backreaction}---Since the string is massive it will curve the spacetime, and this curvature will in turn affect how the string evolves. We study this backreaction effect by decomposing the full metric $g_{\mu\nu}$ into a Minkowski background $\eta_{\mu\nu}$ and a perturbation $h_{\mu\nu}$ due to the string $g_{\mu\nu}=\eta_{\mu\nu}+h_{\mu\nu}$. Crucially, this is a \textit{semi-perturbative} approach: we treat $h_{\mu\nu}$ at linear order and discard all higher-order terms, while treating the evolution of $\epsilon$ and $X^\rho$ non-perturbatively. This is justified because the metric perturbation is of order $G\mu\ll1$ and remains small throughout, whereas changes in the shape and energy of the string accumulate over time.\\
Then, the equation of motion~\eqref{eq:TempOrtho-EoM} becomes
\begin{eqnarray}
    \dv{}{t}\left(\epsilon\dot{X}^\rho\right)-\dv{}{\zeta}\left(\frac{\acute{X}^\rho}{\epsilon}\right)=\mathcal{F}^\rho \label{eq:TempOrtho-Semi-Pert}
\end{eqnarray}
where
\begin{align}
\begin{split}
    \mathcal{F}^\rho=&\left(-\epsilon\dot{X}^\mu\dot{X}^\nu+\frac{1}{\epsilon}\acute{X}^\mu\acute{X}^\nu\right)\eta^{\rho\sigma}\\
    &\quad\times\left(\partial_\mu h_{\sigma\nu} - \frac{1}{2}\partial_\sigma h_{\mu\nu}\right)
    \label{eq:Backreaction-Definition}
\end{split}
\end{align}
Is the gravitational backreaction. The $\rho=0$ equation of~\eqref{eq:TempOrtho-Semi-Pert} tells us that $\dot{\epsilon}=\mathcal{F}^0$, which means that the gravitational backreaction results in a change of energy of the string
\begin{eqnarray}
    \dv{E}{t}=\mu\oint\dd{\zeta} \: \mathcal{F}^0.
\end{eqnarray}
We will see that the net change in energy of the string over a period is negative, so that the string loses energy over time.

To actually compute the gravitational backreaction $\mathcal{F}^\rho$ we follow the gauge-invariant formalism introduced by Chernoff et al.~\cite{Chernoff_2019} and apply it to the temporal-transverse gauge. The metric perturbation $h_{\mu\nu}(x)$ at some point $x$, not necessarily on the string is given by
\begin{eqnarray}
    h_{\mu\nu}(x)=-4G\mu \oint \dd{\zeta}\left[\frac{\sqrt{-\gamma}\:\Sigma_{\mu\nu}}{|r_1|}\right]_{t_{\mathrm{ret}}}, \label{eq:MetricPert1D}
\end{eqnarray}
where $\Sigma_{\mu\nu}=P_{\mu\nu}-\frac{1}{2}g_{\mu\nu} P$, $r_1=\partial_t \sigma$ with the flat-space Synge function $\sigma(x,y)=\frac{1}{2}\eta_{\mu\nu}(x^\mu-y^\mu)(x^\nu-y^\nu)$, and $t_\mathrm{ret}$ is the retarded solution of $\sigma(x,X(t_{\mathrm{ret},\zeta}))=0$.\\
Equation~\eqref{eq:MetricPert1D} can be used to compute the gravitational wave signal emitted by the string. However, to find the gravitational backreaction $\mathcal{F}^\rho$ we need the derivatives $\partial_\rho h_{\mu\nu}$ of the metric perturbation, which are given by
\begin{eqnarray}
    \partial_\rho h_{\mu\nu}(x)=4G\mu\oint\dd{\zeta}{\left[\frac{1}{|r_1|}\partial_\tau\left(\frac{\sqrt{-\gamma}\:\Sigma_{\mu\nu}\Omega_\rho}{r_1}\right)\right]}_{\tau_{\mathrm{ret}}}, \label{eq:Metric-Derivative}
\end{eqnarray}
where $\Omega_\rho = \partial_\rho\sigma(x,X) = x_\rho - X_\rho$ is the 4-vector from the retarded source point to the field point. This result can then be plugged into~\eqref{eq:Backreaction-Definition} to obtain the gravitational backreaction at a point on the string due to the rest of the string, which we will denote by $\mathcal{F}^\rho_\mathrm{int}$.

A key insight in~\cite{Chernoff_2019} was that the derivation of~\eqref{eq:Metric-Derivative}, and subsequently the backreaction, is only valid if the intersection line of the past light-cone and the string worldsheet is smooth. However, there are two scenarios where this generally no longer holds true:
Firstly, the worldsheet is genuinely not smooth at kinks or cusps, which will translate over to the intersection line and introduce boundary terms at these points. Secondly, when taking the field point $x$, at which we evaluate the metric perturbation and its derivative, onto the worldsheet, the intersection line will not be smooth at the field point itself.

In the following we focus on string loops without kinks. Furthermore, since cusps - unlike kinks, which persist throughout the evolution~\cite{Vilenkin:2000jqa} - are transient features, they do not affect the backreaction at most worldsheet points. We therefore ignore the first scenario and retain only the field point contribution. Following the same calculations as in \cite{Chernoff_2019}, but applied to the semi-perturbative temporal-transverse gauge we find that the field point contribution to the backreaction force is given by
\begin{align}
\begin{split}
    &\mathcal{F}^\rho_{\mathrm{field}}=\\
    &4G\mu\left[2\epsilon\ddot{X}^\rho-\frac{\epsilon}{\phi}\left(\partial_t\left(\frac{\phi}{\epsilon}\right)\epsilon\dot{X}^\rho+\partial_\zeta\left(\frac{\phi}{\epsilon}\right)\frac{\acute{X}^\rho}{\epsilon}\right)\right].\label{eq:Field-Point-Force}
\end{split}
\end{align}
The total gravitational force acting on the string is then given by the sum of these two contributions $\mathcal{F}^\rho = \mathcal{F}^\rho_{\mathrm{int}} + \mathcal{F}^\rho_{\mathrm{field}}$. This will be the force that goes into the equations of motion \eqref{eq:TempOrtho-EoM} and is used in the numerical evolution of the string in the following. The details of the numerical implementation of the dynamics can be found in the appendix.

For a smooth Nambu-Goto worldsheet the total gravitational backreaction is finite~\cite{PhysRevD.42.2505,Carter_1998}, a non-trivial result given that the metric perturbation in the vicinity of the field point generally diverges. At a cusp the worldsheet is not smooth, the velocity momentarily reaches the speed of light. and the backreaction in the cusp region is correspondingly large. Nevertheless, the contribution from the cusp region to the total integrated force remains finite~\cite{Chernoff_2019}, so the backreaction problem is well-posed even in the presence of cusps.

Our approach differs from earlier treatments~\cite{PhysRevD.42.2505,Blanco_Pillado_2019,Wachter_2017,Wachter_2017,Wachter_2024} of gravitational backreaction on cosmic string loops in several key respects. First, while previous work accumulates the backreaction over a complete period of oscillation before applying it as a correction to the string trajectory, our formalism evolves the string continuously in time. This allows us to resolve the dynamics \textit{within} a period and, in particular, to track the detailed behaviour near cusps, including the damping of the cusp velocity. Furthermore, the corrected string shape immediately feeds back into the force computation, so that effects like cusp-damping or the shrinking of the string are automatically accounted for. This intra-period resolution also allows us to actually compute the gravitational waveforms produced by the string, since the waveform depends on the detailed dynamics of the string within a period, and not just on the net change after a period. Second, we treat the string as a smooth object throughout, rather than piecewise-linear. A piecewise-linear approximation with finite segment length is not able to accurately model cusp formation. Lastly, the use of the temporal-transverse gauge is consistent with this continuous-time scheme and, as a further benefit, avoids the need to reparametrise the worldsheet after each period, as is required in the conformal gauge~\cite{PhysRevD.42.2505,Blanco_Pillado_2019,Wachter_2024}.\\
All of these improvements are essential to accurately model the effect of the backreaction on the dynamics of the string, and in particular on the behaviour of cusps. Thus our formalism provides a genuine advance in the state of the art of modelling gravitational backreaction on cosmic strings, and allows us to make new predictions for a more realistic string loop, as well as for the gravitational wave signal emitted by such a loop.

\begin{figure}
    \centering
    \includegraphics[width=\linewidth]{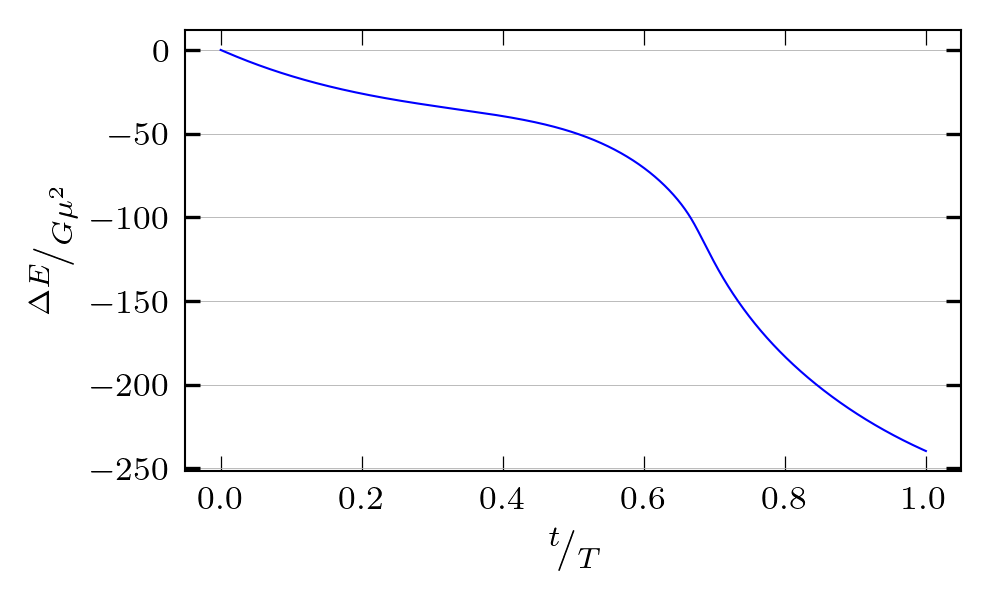}
    \caption{Evolution of the string's energy for the KT loop ($\alpha=0.5$, $\phi=0$). The string tension is $G\mu=10^{-5}$.}
    \label{fig:Energy-Evolution}
\end{figure}

\textit{Results}---All results presented in this section are for the first period of backreacted evolution of an exemplar Kibble Turok loop~\cite{Kibble:1982cb,Turok1984}, i.e.\ the interval from the activation of the backreaction to the end of the first oscillation. However, the results are analogous for other tested loops.\\
The primary effect of the gravitational backreaction is that the string loses energy over time, as shown in Figure~\ref{fig:Energy-Evolution}. We can see that the loss is concentrated around the time of the cusp. We usually describe the energy loss over one period via the numerical coefficient $\Gamma$ defined by
\begin{eqnarray}
    P = - \frac{\Delta E}{\Delta t} = \Gamma G\mu^2.
\end{eqnarray}
Reading off the total change after on period from Figure~\ref{fig:Energy-Evolution} gives $\Gamma\approx 239$, which agrees with the prediction obtained using the quadrupole formula as described in~\cite{Allen_2001} and calculated in~\cite{Chernoff_2019} to within $1\%$. This shows that the energy loss due to gravitational backreaction is consistent with the expected power radiated into gravitational waves. Over longer periods the actual energy loss may diverge from this prediction as the loop shape changes over time.

Furthermore, we find that cusps survive backreaction. The combination $g_{\mu\nu}\dot{X}^\mu\dot{X}^\nu$, where $g_{\mu\nu}$ includes the metric perturbation $h_{\mu\nu}$, still reaches null, and does so for all string tensions. The cusps do however get shifted in time and in space. This agrees with analytic results by Thompson~\cite{PhysRevD.37.283} and previous numerical results~\cite{PhysRevD.42.2505}, which found that cusps survive backreaction.

\begin{figure}
    \centering
    \includegraphics[width=\linewidth]{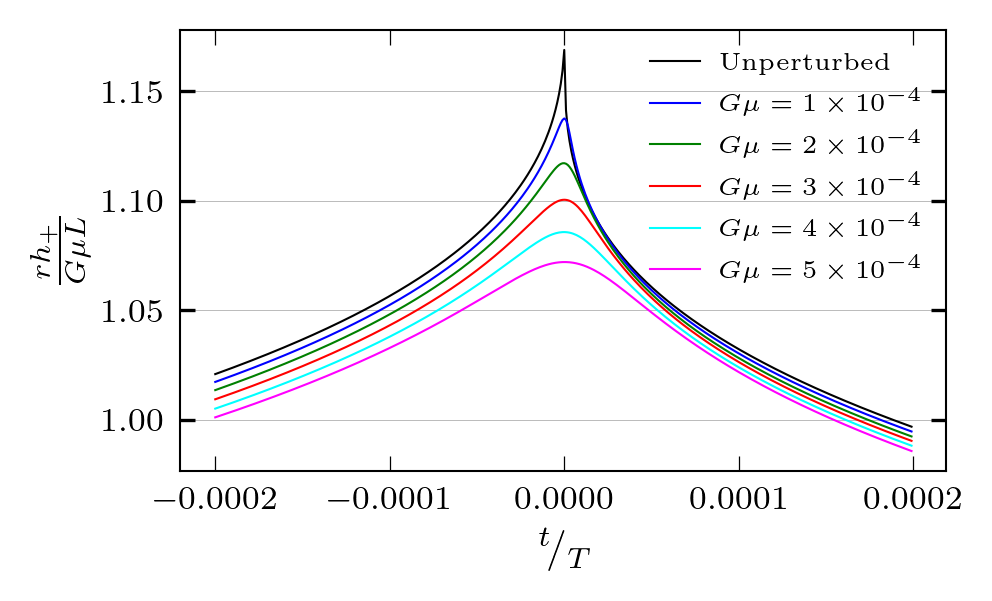}
    \caption{The cusp peak in the gravitational wave signal emitted by a perturbed KT loop with different string tensions. Note that the peaks are aligned in time and rescaled in amplitude for ease of comparison.}
    \label{fig:GW-Signal-Peak-Perturbed}
\end{figure}

Cusps are of particular interest since they emit strong, narrow beams of gravitational waves with a distinct waveform~\cite{Vilenkin:2000jqa,Damour_2001}. 
In the direction of the cusp beam the strain signal has a sharp peak, which can be seen in Figure~\ref{fig:GW-Signal-Peak-Perturbed}.
In fact the peak produced by the cusp is itself a cusp~\cite{Damour_2001}. In frequency space this corresponds to a power law spectrum $\tilde{h}(\omega_n)\propto \omega_n^{-4/3}$, where $\omega_n=\frac{2\pi n}{T}=\frac{4\pi n}{L}$ is the frequency of the $n$-th mode.

\begin{figure}
    \centering
    \includegraphics[width=\linewidth]{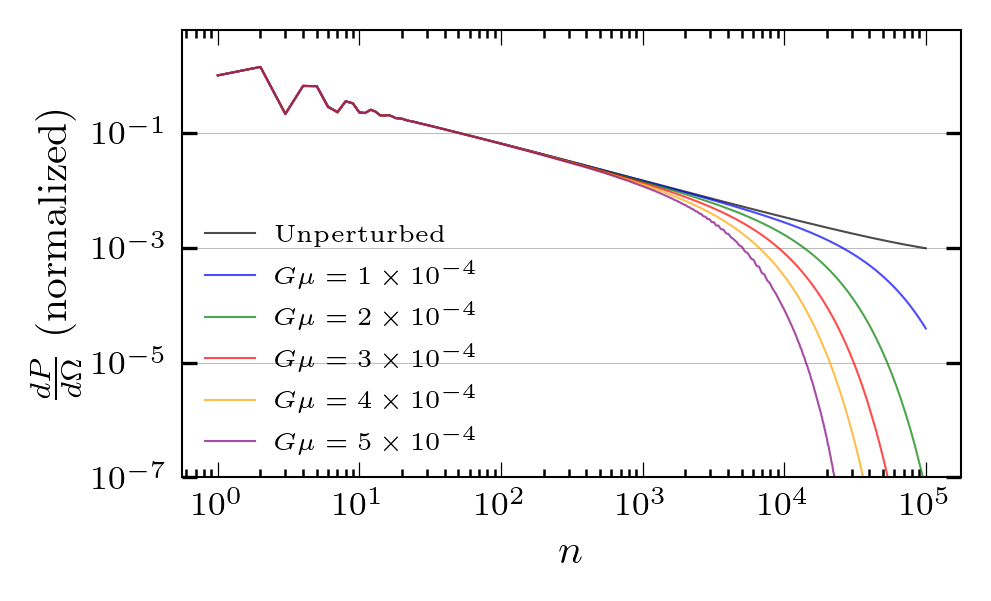}
    \caption{The spectrum of the power radiated in the direction of the cusp beam by a perturbed KT loop with different string tensions, compared to the unperturbed case. Note that the spectra are rescaled in amplitude for better comparability.}
    \label{fig:GW-Signal-Spectrum-Perturbed}
\end{figure}

In the following we characterise the spectrum via the radiated power per solid angle 
\begin{eqnarray}
    \dv{P_n}{\Omega} = \frac{1}{16\pi}\omega_n^2\left(|\tilde{h}_+|^2 + |\tilde{h}_\times|^2\right),
\end{eqnarray}
which combines both polarisations such that the choice of basis does not matter, and is more directly related to the energy loss of the string.
For the unperturbed loop $\dv{P_n}{\Omega}\propto\omega^{-2/3}\propto n^{-2/3}$, as shown by the black curve in Figure~\ref{fig:GW-Signal-Spectrum-Perturbed}. This power law behaviour extends to very high frequencies, up until the string scale~\cite{Damour_2001,Damour_2005}, and well into the frequency bands available to current and near-future gravitational wave detectors. This makes cusps promising targets for gravitational wave searches for cosmic strings~\cite{Damour_2005,Dimitriou_2025}. Note that the power law $\dv{P_n}{\Omega}\propto n^{-2/3}$ implies that the power radiated into the direction of the cusp beam is divergent, however, this singularity is integrable and the total power emitted by the cusp is finite~\cite{Damour_2001,Vachaspati_1985}.

The cusp peak gets modified by gravitational backreaction as shown in Figure~\ref{fig:GW-Signal-Peak-Perturbed}. The sharp peak gets smoothed, with the effect growing with $G\mu$. In frequency space this manifests as a high-frequency cut-off of the $\dv{P}{\Omega}$ spectrum, as can be seen in Figure~\ref{fig:GW-Signal-Spectrum-Perturbed}. The lower frequency modes still follow a power law $\dv{P}{\Omega}\sim \omega^{-2/3}$, while in the high-frequency limit the power spectrum is well described by an exponential $\dv{P}{\Omega}\sim e^{-\omega/\omega_c}$.  Here, $\omega_c$ is the cut-off frequency, determined by fitting the high-frequency spectrum to an exponential, and provides a good estimate of where the cut-off occurs.

\begin{figure}
    \centering
    \includegraphics[width=\linewidth]{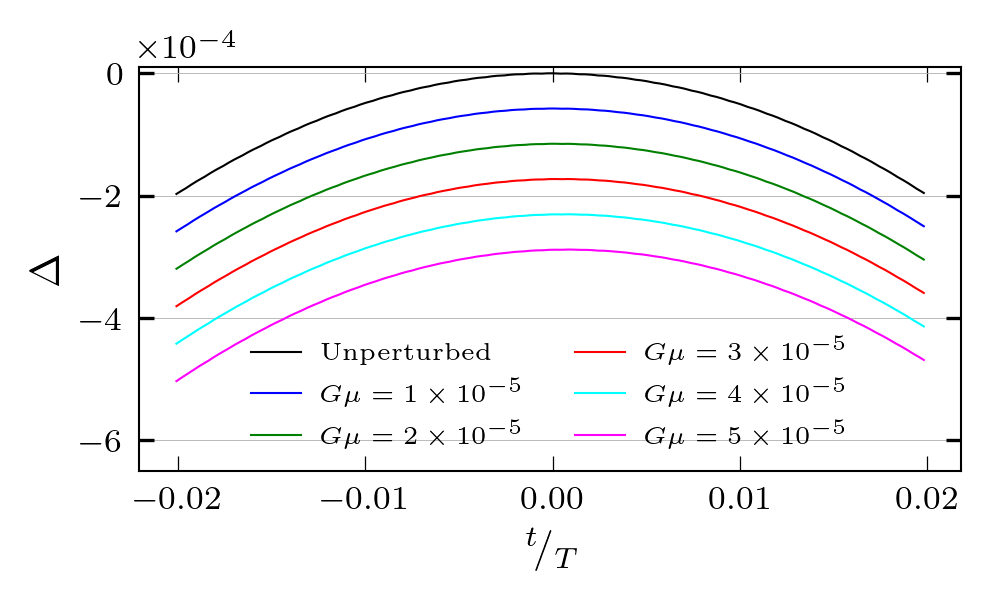}
    \caption{Deviation from the speed of light of the maximum speed along the string near the cusp for the KT loop ($\alpha=0.5$, $\phi=0$) at different string tensions, as seen by a distant observer.}
    \label{fig:Speeds-Tension}
\end{figure}

The reason for this modification is that, although the cusp still locally reaches the speed of light in the perturbed spacetime, its coordinate velocity need not reach unity in the asymptotically inertial frame used to define the gravitational-wave signal (that is, the frame of the far away observer).  Thus, in asymptotically flat coordinates, the emitting string element will have coordinate velocity $\mathbf{v}=\mathbf{\dot{X}}$, with $|\mathbf{v}|<1$. Figure~\ref{fig:Speeds-Tension} shows the corresponding deviation $\Delta=1-|\mathbf{v}|$ of the coordinate speed from the speed of light, which increases with string tension. This asymptotic frame velocity is assigned to the string in the gravitational wave calculation, and its departure from unity produces the observed high-frequency cutoff.

The larger the string tension $G\mu$, the stronger the effect of the backreaction, and so the lower $\omega_c$. Computing $\omega_c$, or $n_c$ to be precise, for different string tensions we find $n_c\propto {(G\mu)}^{-3/2}$. We confirmed this by testing different loop geometries. This scaling can be understood analytically using the scaling arguments of Drew and Rybak~\cite{drew2025newmasslessspectracosmic}. Their analysis models small departures from the exact cusp condition by a dimensionless parameter $\Delta_\pm \ll 1$ quantifying how far left- and right-moving modes at the cusp depart from the unit Kibble-Turok sphere, so that $\Delta_\pm = 0$ corresponds to an exact cusp. Backreaction produces precisely this departure: the modes of the backreacted string do not quite reach unit length, giving $\Delta_\pm > 0$. More precisely, the departure of the modes from the unit sphere due to backreaction is of order $G\mu$, i.e. $\Delta_\pm\sim\Delta\propto G\mu$.

Drew and Rybak (their Eq.~(42)) show that with these perturbations the gravitational wave strain in the direction of the cusp beam takes the form $\tilde{h}(\omega) \sim \omega^{-4/3}\,[\mathrm{Ai}'(\zeta_\pm)]^2,$ where $\zeta_\pm \propto \Delta_\pm\,\omega^{2/3}$. At $\Delta_\pm = 0$, $\mathrm{Ai}'(0)$ is a nonzero constant and so $\tilde{h}\propto\omega^{-4/3}$, recovering the standard Damour-Vilenkin spectrum. For nonzero $\Delta_\pm$ the arguments $\zeta_\pm$ grow with frequency, and in the large-$\omega$ limit $\mathrm{Ai}'(\zeta) \sim e^{-2\zeta^{3/2}/3}$. The strain therefore behaves as $\tilde{h}(\omega) \sim e^{-c\,\Delta^{3/2}\,\omega}$ where $c$ is an $O(1)$ numerical constant absorbing the loop-geometry factors, and so $\dv{P}{\Omega}\sim e^{-2c\,\Delta^{3/2}\,\omega}$. This identifies $n_c\propto\omega_c \sim \Delta^{-3/2} \propto (G\mu)^{-3/2}$, in agreement with our numerical results.\\

For the spectra of the simple loops considered here we find that for string tensions $G\mu\sim10^{-4}$ the spectrum starts being affected by gravitational backreaction around the $n_c\sim10^4$ mode. We can translate this into a cut-off frequency for such loops of arbitrary length and tension via
\begin{eqnarray}
    f_c\sim0.1\frac{c}{L}(G\mu)^{-3/2}\sim\frac{1\mathrm{Hz}}{L_{\mathrm{pc}}}\left(\frac{10^{-6}}{G\mu}\right)^{3/2},
\end{eqnarray}
where $L_{\mathrm{pc}}$ is the length of the string in parsecs. 
For realistic string tensions we have $n_c\gg10^2$, so the cut-off will not affect the stochastic gravitational wave background generated by a network of cosmic string loops, based on the results in~\cite{Caldwell:1996en,Sanidas:2012ee}. However, it can strongly affect searches for transient cusp bursts. Following the calculation of~\cite{Damour_2000}, we find that the cut-off due to backreaction renders these  bursts virtually undetectable in the frequency bands of ground-based detectors such as LIGO and Virgo. In contrast, we expect the bursts to remain detectable with LISA over a range of string tensions bounded as follows: for $G\mu\gtrsim10^{-8}$, the burst is cut off below the LISA band, whereas for $G\mu\lesssim10^{-14}$ the signal is too weak to be detected.

\textit{Conclusion}---We have presented a continuous-time, semi-perturbative backreaction formalism for Nambu-Goto string loops in the temporal-transverse gauge, in which the string trajectory and its gravitational field are evolved simultaneously. Unlike period-by-period approaches, the formalism resolves the intra-period dynamics of the string, enabling, for the first time, direct computation of the gravitational-wave waveform produced by a continuously backreacted loop.

We find that the energy loss of the string due to backreaction is consistent with quadrupole formula estimates. This agreement confirms that the backreaction formalism correctly captures the energy loss of the string due to gravitational wave emission.

A key physical finding is that locally cusps survive backreaction. Nevertheless the gravitational wave signal that reaches a distant observer gets modified: the sharp cusp peak in the waveform is smoothed, and the power spectrum acquires a high-frequency cut-off $f_c\propto(G\mu)^{-3/2}$. We confirm this scaling across distinct loop geometries and interpret this analytically using the Drew-Rybak Airy function analysis~\cite{drew2025newmasslessspectracosmic}, which shows that the exponent is a universal consequence of near-cusp kinematics, while the prefactor varies with loop geometry. The suppression of the gravitational wave signal above $f_c$ reduces the expected signal-to-noise ratio in current and near-future detectors. We find that the cusp bursts are virtually undetectable within the LVK frequency band, but may still be found with LISA, provided that the string tension is in the range $G\mu\sim 10^{-14}-10^{-8}$.

\textit{Acknowledgements}---L.G. s supported by the STFC DiS-CDT scheme and the STFC DTP. Computations were performed using the Swirles (Cambridge) and the DiRAC Cosma (Durham) clusters.

\bibliography{biblio}

\appendix*
\makeatletter
\def\theequation@prefix{A}
\makeatother
\setcounter{equation}{0}

\section{End Matter}


\textit{Appendix: Numerical Implementation}---Following \cite{AllenEvolution}, we introduce the variables $\boldsymbol{\alpha}=\mathbf{\acute{X}}-\epsilon\mathbf{\dot{X}}$ and $\boldsymbol{\beta}=\mathbf{\acute{X}}+\epsilon\mathbf{\dot{X}}$. These are the natural variables for a numerical integration scheme since they allow us to recast \eqref{eq:TempOrtho-EoM} as a system first-order in time evolution equations:
\begin{align}
    \dot{\epsilon} &= \mathcal{F}^0 \label{eq:Epsilon-EoM}\\
    \boldsymbol{\dot{\alpha}} &= -\left(\frac{\boldsymbol{\alpha}}{\epsilon}\right)'-\boldsymbol{\mathcal{F}} - \lambda_\alpha \mathcal{C}_\alpha \boldsymbol{\alpha} \label{eq:Alpha-3-EoM} \\
    \boldsymbol{\dot{\beta}} &= \left(\frac{\boldsymbol{\beta}}{\epsilon}\right)'+\boldsymbol{\mathcal{F}} - \lambda_\beta \mathcal{C}_\beta\boldsymbol{\beta}\label{eq:Beta-3-EoM}\\
    \mathbf{\dot{X}} &= \frac{1}{2\epsilon}\left(\boldsymbol{\beta}-\boldsymbol{\alpha}\right), \label{eq:Vel-3-EoM}
\end{align}
The additional terms $\mathcal{C}_\alpha=g_{\mu\nu}\alpha^\mu\alpha^\nu$ and $\mathcal{C}_\beta=g_{\mu\nu}\beta^\mu\beta^\nu$ with $\alpha^\mu=(-\epsilon,\boldsymbol{\alpha})$ and $\beta^\mu=(\epsilon,\boldsymbol{\beta})$ act as constraint damping terms, with $\lambda_\alpha$ and $\lambda_\beta$ real damping coefficients. From the temporal-transverse-gauge conditions $\mathcal{C}_\alpha$ and $\mathcal{C}_\beta$ vanish analytically in the continuum, but on a discrete grid numerical errors cause them to drift, thus necessitating the damping.\\
Equations~\eqref{eq:Epsilon-EoM}--\eqref{eq:Vel-3-EoM} are integrated using a fourth-order Runge-Kutta scheme with a five-point stencil for spatial derivatives. During the evolution we track the constraints $\mathcal{C}_\alpha$ and $\mathcal{C}_\beta$ as our diagnostic variables, since they are directly related to the gauge conditions. 

The simulation begins with the loop evolving for one full period in flat space without backreaction, establishing a clean initial configuration of $\epsilon$, $\boldsymbol{\alpha}$, $\boldsymbol{\beta}$, and $\mathbf{X}$. The gravitational backreaction and metric perturbation are then activated. At this point $\boldsymbol{\alpha}$, $\boldsymbol{\beta}$ and $\epsilon$ must be adjusted to satisfy the null constraints with respect to the perturbed metric; this adjustment also shifts the total energy of the string, which we interpret as the effect of embedding the string in the curved background. To avoid numerical artefacts the backreaction is not activated at the moment of the cusp, but at a safe time interval afterwards.
\end{document}